\documentclass[singlecolumn,showpacs,showkeywords,prb]{revtex4-1}
\usepackage{graphicx}
\usepackage{dcolumn}
\usepackage{epstopdf}
\usepackage{amsmath}
\usepackage{latexsym}
\newcommand{\ptl}{\partial}
\newcommand{\be}{\begin{equation}}
\newcommand{\ee}{\end{equation}}
\newcommand{\bqr}{\begin{eqnarray}}
\newcommand{\eqr}{\end{eqnarray}}
\newcommand{\half}{{1\over 2}}
\newcommand{\met}{{\sqrt -g}}
\newcommand{\pr}{\prime}
\begin{document}
\title{Cosmological Correlations in Power Law Inflation models}

\author{Amit Kundu}
\email{amit_kundu@physics.iiests.ac.in}
\affiliation{Department of Physics, Indian Institute of Engineering Science and Technology, Shibpur, India}

\author{Prasenjit Paul}
\email{prasenjit071083@gmail.com}
\affiliation{Department of Physics, Govt. College of Engineering and Ceramic Technology, Kolkata, India}


\begin{abstract}

Scalar field with non-minimal coupling to curvature scalar is studied in Robertson-Walker background. The infrared limit of two point function, and, in turn, of the energy-momentum tensor of scalar field have been considered in the power law inflation model. In this limit, within the perfect fluid model, consistent evolution of scale factor following power law inflation gives rise to growing mode solution for negative value of coupling constant. A simplified calculation for density perturbation in power law inflationary models is presented with these mode functions. Salient features of the perturbation spectra has been commented upon.

\end{abstract}

\maketitle

\section{Introduction}

Apart from solving some of the major problems of standard cosmology, inflationary models provide the most elegant solution to the formation of large scale structure in the Universe. The small inhomogeneity in density, as observed in the present day Universe, is considered to be originated from the quantum fluctuation in energy density of scalar field within a small causal region at the early epoch which subsequently grew into super horizon size, leaving horizon during rapidly expanding inflationary phase of the Universe. 
These super horizon size perturbations on re-entering the horizon at the post inflationary phase leaves imprint of primordial fluctuation on the cosmic microwave background radiation (CMBR). The density perturbation spectrum thus generated is nearly scale-invariant and the observable deviation from this scale invariant nature can discriminate among inflationary  models.

The evaluation of the density perturbation depends on the time evolution of scalar field in the time-dependent cosmological background described by Robertson-Walker metric. The observed perturbation being very small \cite{Dunkley}, closely Gaussian, one may feel that free scalar field theory can adequately describe this fluctuation given two point correlation, thereby determining the necessary power spectrum, and all higher even correlations are factorized in terms of them. Moreover, the scalar field are usually considered to be ultra-light ($m<< H$) as the long wavelength limit is truly important in determining perturbation spectrum.

Recent times, the scope of precision observation of CMBR has generated considerable interest to search for its non-Gaussian nature \cite{Bartolo}, albeit very small, following the seminal work of Maldacena\cite{Maldacena}. Within the single field inflationary model, an estimate of the non-Gaussian nature was given for three point correlations involving different numbers of scalar and  gravitational fields to first order in their interactions.
Later on, Weinberg \cite{Weinberg, Weinberg2} has refined the formulation using closed time path (`in-in formalism') and has shown a general method of calculation of cosmological correlations to higher orders in perturbation theory. Subsequently,  Weinberg also pointed out some of the pertinent issues related to infrared divergence and the scheme of loop calculations, in general\cite{Weinberg}. Some of the works employing `in-in formalism' can be mentioned in this context \cite{Lim and Others}.

The non-Gaussian distributions can be obtained in many ways such as going beyond models of single-field inflation \cite{Lyth-Jokinen-Mazumdar}, using extended kinetic terms \cite{Langlois-Steer-Arroja-Tanaka}, or by breaking the slow-roll conditions in single-field inflation\cite{Chen-Lim}. This has been comprehensively reviewed \cite{Bartolo}.
  
It is worth mentioning that the scalar field dynamics in time-dependent Robertson -Walker background has been studied earlier using real time formalism  starting both from initial
condition in equilibrium \cite{Semenoff-Weiss1} or in non-equilibrium \cite{Jordan-Hu} with techniques similar to flat space-time \cite{Schwinger, Landsman}. Typically this involves evolution of field in complex time domain, resulting in multiple Green functions which may be used to develop a consistent scheme of perturbation theory. In particular, considerable interest has been shown towards de Sitter space-time which has cosmological relevance. 

Usually, most inflationary models employs de Sitter phase of expansion with horizon staying fixed but physical wavelength growing  and the nature of potential driving such inflation should obey slow roll conditions to generate enough inflation to cover present horizon. The power law evolution of scale factor $a(t)\sim t^p$, with $p$ large, has been shown \cite{Matareese} to be adequate in solving almost all the problems that are addressed by the usual models of inflation. The power law inflation has been studied in the context of extended inflation\cite{Guth-Jain-Kolb etal}. Further works \cite{Sahni 1,Sahni 2, Liddle} in this context are worth mentioning . 
The power law inflation are mostly associated with canonical scalar field with exponential potential. It may also happen otherwise, such as in K-inflation model\cite{Garriga}.
In view of recent available data, such as Planck, the feasibility of power law inflation can be explored in several fronts.

Scalar field with nonminimal coupling to curvature scalar has generated considerable interest to find a viable model of inflation and density 
fluctuation consistent with observation even within the limit of very weak self-coupling has been explored \cite{Unruh papers}. 
In this work we have considered scalar field coupled to gravity with arbitrary coupling to scalar curvature using real time formulation of field theory. This formulation has the advantage
of clear separation of the effect of initial condition from the dynamics of background. 
 For ultra-light scalars, the mode functions are obtained in the Robertson-Walker background with scale factor grows according to power law. 
 The expectation value of energy-momentum tensor operator in the infrared limit is considered perfect fluid model in the spirit of Ford-Parker\cite{Ford-Parker}. As the infrared modes are relevant for structure formation at large scale afterwards, we address the issue of calculation of density perturbation in power inflation with real time formulation a la Semenoff-Weiss \cite{Semenoff-Weiss2}. The time evolution is followed using density operator at the onset of inflation which mimics 
equilibrium form with parametrization by $\beta_0$. The correlation of density fluctuation is obtained by restricting up to quadratic order term to see small effect of nonlinearity. Finally, the dependence of spectral index on nonminimal coupling and power of scale factor has been pointed.

In section 2 basic features of scalar field theory with non-minimal coupling to curvature scalar are stated. The mode function in the background of Robertson-Walker is given and its relevant asymptotic form is utilized to obtain the form of energy-momentum tensor in the infrared limit. Particularly, the consistent picture of fluid model has been identified to have its relevance in the context of power law inflation. Section 3, contains calculation of density perturbation, defined ab initio, and its dependence on two point correlations has been mentioned \cite{Mallik-Raichaudhuri}. The result of gauge invariant perturbation theory has been employed to bridge perturbation at the `horizon exit' during inflation to its reentry in post-inflationary radiation or matter dominated phase. Section 4 contains a sample calculation of density perturbation in power law inflation model employing real time path formalism. Some of the features of the perturbation spectrum thus obtained has been discussed.

\section{Scalar field in Robertson-Walker background}

The action for a real scalar field $\phi$ in a general space-time background with non-minimal coupling to scalar curvature ${\cal R}(t)$ is given by
\be  S=\int d^4x{\sqrt -g}\bigg[\frac{1}{2} g^{\mu\nu}{\nabla_\mu}{\Phi}(x){\nabla_\nu}{\Phi}(x)- \frac{1}{2}\big\{m^2+\xi{\cal R}(t)\big\}\bigg]\Phi^2, \label{eq:action}\ee \\
where metric $g^{\mu\nu}$ is the space-time metric and $\xi$ denotes the coupling constant, taken arbitrary.
The coupling constant $\xi =0$ corresponds to minimally coupled scalar field while $\xi = {1\over 6}$ corresponds to conformally coupled scalar field.
The field equation obtained by varying the action is given by \be \left[{1\over\met}\ptl_\mu(g^{\mu\nu}\met \ptl_\nu\phi) +m^2 +\xi {\cal R}(t) \right] \phi(x)=0.\label{eq:field equation}\ee
Considering Robertson-Walker metric background for spatially flat section 
\[ ds^2=dt^2 -a^2(t)[{dr^2}+r^2(d\theta^2+\sin^2\theta~ d\varphi^2)],\]the above equation leads to
\be {\ddot\phi} + 3H{\dot\phi} -{1\over a^2} \nabla^{2}
\phi +(m^2+\xi {\cal R}(t))\phi =0. \ee
The field can be Fourier decomposed into modes 
 \be \phi({\bf x})=\int d^3k [a_k e^{i{\bf k}\cdot{\bf x}}f_k(t)+a^\dagger_k e^{-i{\bf k}\cdot{\bf x}} f_k^\star(t)]. \ee
The time dependent mode function, $f_k(t)$ obeys
\be  \left[ {d^2\over dt^2}+3{{\dot a }\over a }{d\over dt} +{k^2\over a^2(t)}+m^2+\xi{\cal R}(t)\right] f_k(t)=0.\ee
Writing $f_k(t) =a^{-3/2} g_k(t)$, the above equation reduces to that of harmonic oscillator with time dependent frequency
\be {\ddot g}_k(t) +\omega_k^2(t) g_k(t) =0,\label{eq:mode equation}\ee where
\be \omega^2_k(t)= \left({ k^2\over a^2(t)}+m^2\right) +(\xi -1/6){\cal R}(t)-\half\Big[{\dot H(t)} +\half H^2(t) \Big].\ee
Note that, the terms within the first parenthesis do not involve any time derivative of scale factor while the
remaining terms are comprised of such time derivatives, at most to second order.

The Wronskian of complex mode functions obey
\be {\dot f}_kf_k^\star -f_k{\dot f_k}^\star =-i/a^3(t)~~,\ee
which may be chosen so as to satisfy at the initial time, say $t_0$. Written in terms
of $g_k$, the Wronskian condition reads
\be {\dot g}_kg_k^\star -g_k{\dot g_k}^\star =-i,\label{eq:wronskian}\ee
which is time independent and therefore it is satisfied for all subsequent times.
It is obvious that the detailed form of mode function $g_k(t)$ depends on the time dependence of the scale factor.
We are primarily interested in cases where the scale factor either grows exponentially with time, as in de-Sitter
phase, or it follows a simple power law dependence on time.

For power law behaviour of the scale factor, $a(t)\sim t^p$, the mode equation~(\ref{eq:mode equation}) reduces to
\be {\ddot g}_k(t) +\Big[{k^2\over a^2(t)}+H^2(t)\Big\{{m^2\over H^2(t)}+12\xi\Big(1-{2\over p}\Big)
- {9\over 4}\left(1-{2\over 3p}\right)\Big\}
\Big]g_k(t) =0. \ee

The mode equations are solved in general using WKB approximation.
However, in the limit where mass term $m^2/H^2(t)$ is either vanishingly small as for ultra light scalar
($m<<H$) or time independent, such as in effective mass description, the mode functions obeying Wronskian (\ref{eq:wronskian}) can be obtained exactly
\be g_k(t) = \half {\sqrt {{p\over p-1} {\pi\over H}}} H_\nu^{(1,2)} \left(
{p\over p-1}{k\over a(t)H(t)}\right), \label{eq:value of mode function}\ee
where $H_\nu^{(1,2)}$ are the Hankel functions of first and second kind respectively with order $\nu$
with \be \nu^2 = {(3p-1)^2\over 4(p-1)^2} - 6\xi {p(2p-1)\over (p-1)^2}-\left({p\over p-1}\right)\left({m^2(t)\over H^2(t)}\right). \ee

Incidentally, in the de-Sitter phase $a(t)\sim e^{H_0t}$, $H_0$ being a constant,
the corresponding equation is
\be {\ddot g}_k(t) +\left[\left({ m^2\over H_0^2} +12\xi-{9\over 4}\right)H_0^2+{k^2\over a^2(0)}e^{-2H_0t}\right] g_k(t) =0 ,\ee
and the solution is given by
\be g_k(t) = \half {\sqrt {\pi\over H_0}}{H_\nu ^{(1,2)}}\left({k\over a(t)H_0}\right), \ee 
with $ \nu^2 ={9/4}-12\xi -{m^2/H_0^2}$.

Often it is convenient to use conformal time $\eta=\int dt/a(t)\,(-\infty <\eta <0)$.
In terms of conformal time field decomposed in modes 
\be \phi({\bf x})=\int d^3k [a_k e^{i{\bf k}\cdot{\bf x}}f_k(\eta)+ a^\dagger_k e^{-i{\bf k}\cdot{\bf x}} f_k^\star(\eta^\pr)]. \label{eq:mode expansion}\ee
The mode functions written in general,
\be f_k(\eta) = \alpha\left({\pi\eta\over 4}\right)^{1/2}\eta H \Big[C_1H^{(1)}_\nu(k\eta)+C_2H_\nu^{(2)}(k\eta)\Big], \ee 
where $\alpha=~1~{\rm{or}}~\mid{p-1\over p}\mid$ for de Sitter or power law behaviour of scale factor, respectively. 
Here the coefficients $C_1,C_2$ may be chosen as $k$ dependent functions. 
The Wronskian condition (\ref{eq:wronskian}) imposes relation between coefficients; $\mid C_2\mid^2 - \mid C_1\mid^2=1$.

The vacuum state is defined as \be a_{\bf k}\mid 0\rangle =0, \ee where $a_{\bf k}$ is the annihilation operator of the field quanta.

In contrast to Minkowski space field theory where unique choice of vacuum is guided by Poincar${\acute e}$ invariance, such choice of vacuum
is absent in curved space-time. Different choices of constants $C_1$ and $C_2$ obeying above condition leads to plethora of 
inequivalent vacua, all related by Bogoliubov transformation. This may also exhibit infrared divergence
even in the absence of interaction, except gravity, as shown in Robertson-Walker background \cite{Ford-Parker}. Importantly, based on the
Hadamard structure of two point correlation, in general background, a state exhibiting 
infrared divergence cannot arise from dynamical evolution from regular initial condition \cite{Wald, Fulling et al}.

The two point correlation is given by
\be \langle\phi(x)\phi(x^\pr)\rangle = \int {d^3k\over (2\pi)^3}e^{i{\bf k}\cdot ({\bf x}-{\bf x}^\pr)} f_k(\eta)f_k^\star(\eta^\pr). \ee

Here we are interested in the problems of density fluctuation involving modes of long wavelength, typically of the order of Hubble radius.
The ultraviolet divergence associated with high energy modes and its renormalization has been investigated thoroughly \cite{Birrel-Davies},
which is hardly a practicable issue from the perspective of calculating density perturbation \cite{Lyth, Bartolo2}.

The infrared contribution of the above integral can be obtained by employing expansion of Hankel function for small argument
($\nu > 0$ and up to first non leading term \cite{Abramowitz};
\be H_\nu^{(1,2)}(z)= \mp {i\over \pi}\Gamma(\nu)\left({z\over 2}\right)^{-\nu}
\Big[1+{1\over \nu-1}\left({z\over 2}\right)^2\Big]   \label{eq:Hankel function} \ee 
giving mode function
\be f_k(\eta)~\stackrel{k\rightarrow 0}{\sim}~ (C_2 -C_1)A\eta^{\mu -\nu}k^{-\nu}[1+\gamma k^2\eta^2],\ee
where $\mu$ =$(3p-1)/[2(p-1)]$,~ A=$\pm{i\over \pi}\Gamma(\nu)2^{\nu-1}(1-p)^{p/(p-1)}$ ~ ${\rm{and}}~~ \gamma={1/(\nu-1)}$.

Clearly, the source of divergence is the integral in the infrared limit
\be \mid A\mid^2 \Gamma^2(\nu)\left({\eta\eta^\pr \over 4}\right)^{\mu-\nu}\int d^3k k^{-2\nu}\mid C_2-C_1\mid^2. \ee
Assuming the coefficients are regular as $k\rightarrow 0$, the integral diverges for $\nu \geq 3/2$ .

One may note that similar marginal divergence occurs for a minimally coupled massless scalar field in 
de Sitter phase \cite{Allen-Folacci,Ford-Vilenkin}.

\begin{figure}
	\includegraphics[scale=0.5]{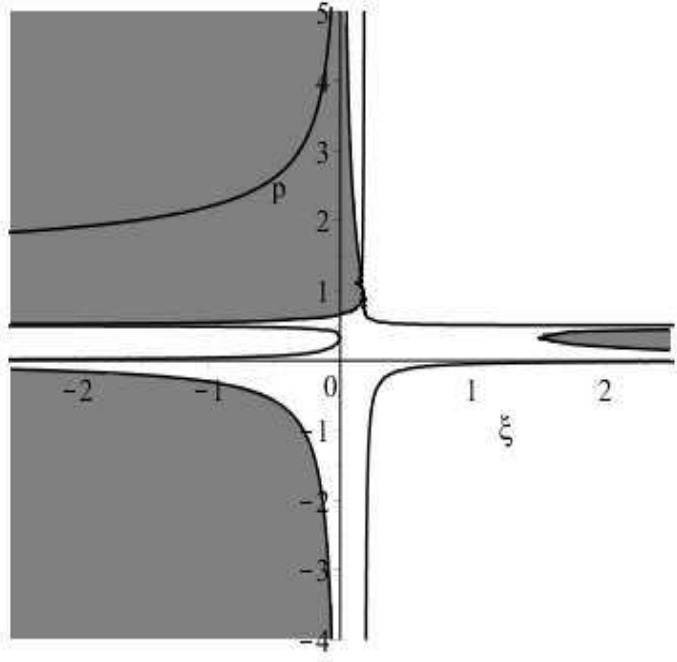}
	\caption{IR divergent regions (shaded) of two point function.}
	\label{fig:condition}
\end{figure}

A plot of such condition \cite{Sahni-Habib}, is shown in ($\xi, p$) plane (figure~\ref{fig:condition}), wherein the shaded region indicates
divergent two point function, in general. However, the coincident limit indicated there is hardly of practical interest which will be pointed later.

For power law expansion, with regular coefficients the infrared divergence in two point function occurs over a wide range of values
of $p$, governed by the inequality $$\xi\leq {3p-2\over 6p(2p-1)}.$$ 
This also shows that for minimally coupled scalar field $(\xi=0)$, 
the divergence occurs for $2/3\leq p< 1$ and $p>1$. The region $\xi < 0$, is interesting as we show presently that growing mode solution occurs in that region.

Further, the infrared divergence can be ameliorated with suitable choice of mode function such that $\mid C_2-C_1\mid$ increases at a rate faster than
$k^{2\nu -2}$. The $k$ dependent coefficients can be determined such as in radiation dominated pre-inflationary phase\cite{Ford-Vilenkin,Sahni-Habib}.\\


The classical energy momentum tensor of a non minimally coupled scalar field can be obtained by varying action (\ref{eq:action}) with respect to the metric $g^{\mu\nu}$:
\bqr T_{\mu\nu}= (1-2\xi)\ptl_\mu\phi\ptl_\nu\phi -\half(1-4\xi)g_{\mu\nu}\ptl_\alpha\phi\ptl^\alpha\phi - 2\xi\phi\nabla_\mu\nabla_\nu\phi +2\xi g_{\mu\nu}
\phi\Box\phi - \xi({\cal R}_{\mu\nu}-\half g_{\mu\nu}{\cal R})\phi^2 +\half m^2g_{\mu\nu}\phi^2. \label{eq:energymomentum tensor}\eqr

The trace of energy momentum tensor is given by
\bqr T\equiv g^{\mu\nu}T_{\mu\nu}= (6\xi -1)\nabla_\mu(\phi\nabla^\mu\phi)
+m^2\phi^2 +\phi(\Box +\xi{\cal R}+m^2)\phi, \label{eq:trace}\eqr 


where the last term vanishes by equation of motion (\ref{eq:field equation}).

Treating $\phi$ as the full quantum field with $\langle\phi\rangle=0$, the energy density and the homogeneous pressure are given by
\be \rho= \langle T_{00} \rangle~,\qquad p=-{1\over a^2}\langle T_{ii}\rangle \qquad
{\rm{(no~sum~over~{\it{i}})}},\ee
and considering fluid model, one may write
\be w={1\over 3}\left(1-{\langle T\rangle\over \langle T_{00}\rangle}\right) \label{eq:omega}.\ee

In spatially flat Robertson-Walker metric,  the classical energy density (\ref{eq:energymomentum tensor}) becomes,

\be T_{00}=\half{\dot\phi}^2 + \half\left({\nabla\phi\over a}\right)^2
+ 6\xi H\phi{\dot\phi} -{2\xi\over a^2}\Big[(\nabla\phi)^2 -\phi \nabla^2\phi\Big] +{\half m^2\phi^2 +3\xi H^2\phi^2}, \ee
and the trace of energy-momentum tensor (\ref{eq:trace}) is given by
\be T=(6\xi -1)\Big[{\dot\phi}^2-\left({\nabla\phi\over a}\right)^2
+\phi{\ddot\phi}+ 3H\phi{\dot\phi} -\phi\left({\nabla^2\phi\over a^2}
\right)\Big] +m^2\phi^2 .~ \label{eq:trace mode}\ee

Using mode expansion (\ref{eq:mode expansion}) of field, the vacuum expectation value of (symmetrized) $T_{00}$ is
obtained;

\be \rho = \int {d^3k\over (2\pi)^3}
\Big[ {1\over 2a^2}\mid f_k^\pr\mid^2 +\half\left(m^2+{k^2\over a^2}+3H^2\xi
\right) \mid f_k\mid^2 +3 \xi{H\over a}(f_k f_k^{\star\pr}+
f_k^\pr f_k^\star) \Big] . \ee

Similarly the trace of energy momentum tensor (\ref{eq:trace mode}) obtained;
\be \langle T\rangle =\int {d^3k\over (2\pi)^3}\Big[ (6\xi -1)
\Big({1\over a^2}\mid f_k^\pr\mid^2 - \omega_k^2\mid f_k\mid^2 \Big ) +m^2\mid f_k\mid^2 \Big] , \ee
where $\omega_k^2 = k^2/a^2(t)+m^2+\xi{\cal R}$ and prime denotes derivative
with respect to conformal time $\eta$.
In the massless limit, the long wavelength behaviour of the energy density and the trace of energy-momentum tensor can be evaluated
using approximation (\ref{eq:Hankel function}). 


\noindent For $\nu\not=\mu,$

\be \rho~ \stackrel{k\rightarrow 0}{\sim}~\mid C_2-C_1\mid^2 \mid A\mid^2 \eta^{-2(\nu-\mu)}{1\over 2\pi^2a^2\eta^2}
\Big\{\half(\nu-\mu)^2 + {3\over 4}\xi(2\mu-1)^2+{3\over 2}(2\mu-1)(\nu-\mu)
\Big\}\int dk k^{2-2\nu} \label{eq:energydensity}\ee
\be \langle T\rangle ~\stackrel{k\rightarrow 0}{\sim}~ \mid C_2-C_1\mid^2\mid A\mid^2 \eta^{-2(\nu -\mu)}{(6\xi -1)\over 2\pi^2a^2\eta^2} \Big\{(\nu-\mu)^2 -{3\over 2}(4\mu^2-1)\xi \Big\}\int dk k^{2-2\nu}.\label{eq:pressure}\ee

For $\nu=\mu$ (with $\mu>0$):
\be \rho~ \stackrel{k\rightarrow 0}{\sim}~\mid C_2-C_1\mid^2 \mid A\mid^2 \Big\{~{3\over 2}~
{\xi\over \pi^2a^2\eta^2}\int dk k^{2-2\mu}+{1\over 4\pi^2a^2}\left(1-6\xi{2\mu -1\over \mu-1}\right)\int dk k^{4-2\mu}~\Big\},\ee
\be \langle T\rangle~ \stackrel{k\rightarrow 0}
{\sim}~\mid C_2-C_1\mid^2 \mid A\mid^2 \Big\{~{3\over 4}~{\xi
	(6\xi -1)\over \pi^2a^2\eta^2} \int dk k^{2-2\mu} - {6\xi -1\over 2\pi^2a^2}\int dk k^{4-2\mu}~\Big\}.\ee


Now, for the consistency of Friedmann equations, with vanishing cosmological constant, the power law behaviour of scale factor requires $w=2/3p-1$ where
$w$ contains the ratio which is considered to be well behaved in the long wavelength limit.

Using eqn(\ref{eq:energydensity}),(\ref{eq:pressure}) we can have from eqn(\ref{eq:omega}) the explicit expression of w as,
\be w = \frac{1}{3}\left[1-\frac{\left\{{({\mu}-{\nu})}^2- \frac{3}{2}(4{\mu}^2-1)\right\}(6{\xi}-1)}{\frac{1}{2}({\mu}-{\nu})^2+ \frac{3\xi}{2}(2\mu-1)^2+\frac{3}{2}(2\mu-1)({\mu}-{\nu})}\right] \label{eq:omega1}.\ee
Also the power law behaviour needs
\be {\langle T\rangle\over \langle T_{00}\rangle} = 4-2/p. \ee

Combining together, for $\nu \not=\mu$, one arrives at the condition,
\be {8(6\xi-1)(\nu-\mu)\nu\over 2(\nu-\mu)^2 +6\xi(\nu-\mu)(2\mu -1)+3\xi (2\mu-1)^2} = {2(2\mu +1)\over 2\mu -1}.\label{eqn:consistent condition}\ee

Next we checked for a consistent solution obeying the constraint, $ \nu^2 = \mu^2 -{3\over 2}(4\mu^2-1)\geq 9/4 $ for growing modes.
This is indeed possible as shown by solid contour falling in shaded region of figure (\ref{fig:condition}). Thus it is possible to have modes leading to growing 
structure in the power law inflation models and it happens for $\xi < 0$. Also a mode convergent initially always remains so in course of time evolution. 
It is apparent from Fig.(\ref{fig:condition}) that for conformal coupling $\xi= 1/6$, one has well behaved two point function for any value of $p >1$.


\begin{figure}[tbp]
	\centering
      \includegraphics[scale=0.5]{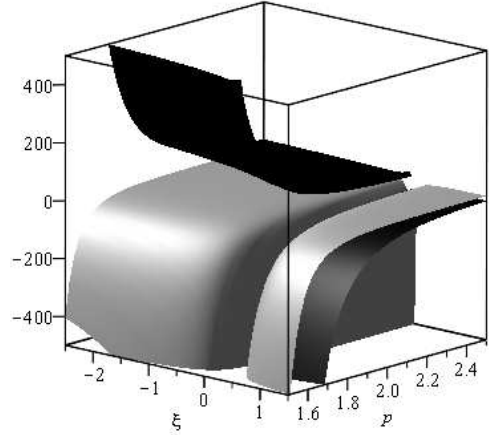}     
	\caption[energydensity domain]{\small  Long wavelength behaviour of energy-density (gray) and pressure (black).}
	\label{fig:energydensity}
\end{figure}

Thus the energy density of non minimally coupled scalar field shows infrared divergence in the same range $(\nu\geq 3/2)$ as the two point
function. However, for $\xi =0$, the energy density of massless scalar is
divergent for $\nu\geq 5/2$ ${\it{i.e.}}~ 3/4\leq p \leq 2,~(p\not=1),$ which is smaller than the corresponding range for two point function\cite{Ford-Parker}.
It may be noted that in figure~(\ref{fig:energydensity}), in this region  $\rho$ and $p$ differs in sign, as expected in inflation.

\begin{figure}[tbp]
		\centering
        \includegraphics[scale=0.5]{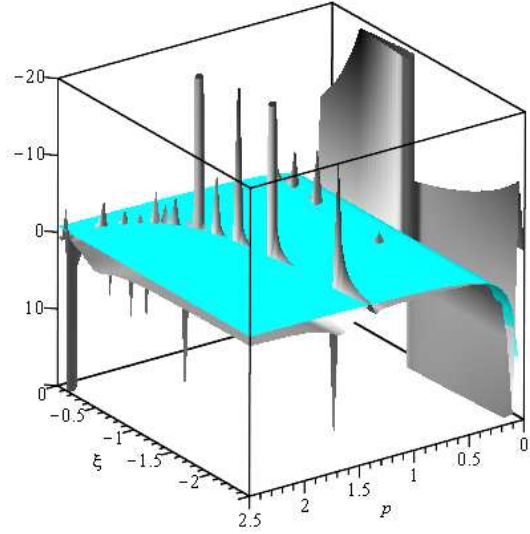}              
		\caption{Distribution of equation of state $w$}
			\label{fig:omega}
	\end{figure}

For $\nu=\mu$, which essentially corresponds to $\xi =0$ (as $\nu >0$), one obtains $w=-1/3$. The solution of Friedmann equation 
and also figure (\ref{fig:omega}) gives $p=1,$ which is in contradiction with divergence range stated earlier \cite{Ford-Parker}.

	Figure~\ref{fig:omega} shows the distribution of $w$. The gray part represents the right hand side of equation (\ref{eq:omega1}) 
	and the coloured part is $w=2/3p -1$ which intersect along a curve.
	Apart from several divergent spikes at follows the same trend as shown by solid line in figure ({\ref{fig:condition}}).
	
	Summarily, for non minimally coupled scalar field, the infrared behaviour of states with $C_1$ and $C_2$ regular is such that 
	metric with power law expansion may have self consistent solutions of Friedmann's equation but it admits infrared divergence, as shown for a minimally coupled scalar \cite{Ford-Parker}.
	In other words, inflation needs choice of special kind states with $C_1$ and $C_2$ irregular and $\xi$ negative.


\section{Density fluctuation in inflationary epoch}

Considering a large volume V at cosmological scale, the density inhomogeneity, at time $t$ is measured by the mean square fluctuation
in the density function \be \left({\delta	\rho\over \rho}\right)^2 =~\left\langle \left({\rho({\vec x},t)-
{\bar\rho}(t)\over {\bar \rho}(t)}\right)^2\right\rangle_x ,\label{eq:density fluctuation}\ee where $\langle\cdots\rangle_x$
denotes average over all space and ${\bar\rho}$ is the homogeneous background energy density,
${\bar\rho} = \langle \rho ({\vec x},t)\rangle_x$.
The density contrast $\delta ({\vec x},t)$, excess over the averaged energy density $\bar{\rho}(t)$, can be Fourier decomposed as
\be \delta({\vec x},t)={1\over \sqrt V}\sum_k\delta_k(t)e^{i{\vec k}.{\vec x}}. \ee

The mean square fluctuation (\ref{eq:density fluctuation}) can be written in terms of Fourier modes
\be \left({\delta\rho\over \rho}\right)^2 = {1\over V}\sum_k\mid\delta_k(t)\mid^2 \longrightarrow \int d(ln~k) {k^3\over 2\pi^2}\mid\delta_k(t)\mid^2. \ee
A convenient measure of fluctuation is the power per logarithmic interval in wave number or variance
\be \left({\delta\rho\over \rho}\right)_k^2 = {k^3\over 2\pi^2}\mid\delta_k(t)
\mid^2~~\equiv \triangle^2(k). \ee
As the density fluctuation arises from quantum fluctuation of field in the inflationary phase, one may obtain similar expression for density inhomogeneity,
with $\rho({\vec x}, t)$ replaced by the corresponding energy density operator $\hat{\rho}({\vec x}, t)$ with expectation
value in an appropriate state. However, the energy density operators involves product of fields at the same space-time
point which is ill defined. This problem may be circumvented by introducing smeared density function operator in a manner similar to the
classical situation.\cite{Mallik-Raichaudhuri}

The energy density operator ${\hat\rho}({\vec x},t)$ is the time-time component
of the energy-momentum tensor of inflaton field:
\be T_{\mu\nu}=\partial_\mu\Phi\partial_\nu\Phi-g_{\mu\nu}
\left({1\over 2}g^{\alpha\beta}\partial_\alpha\Phi\partial_\beta\Phi -V(\Phi)\right),\label{eq:rho fluctuation}\ee
where the potential function $V(\Phi)$ depends on the model chosen. 
With a shift of the field $\Phi(x)$ by homogeneous classical field $\phi(t)$,
\[\Phi({\vec x}, t) = \psi(t) +\phi({\vec x},t),\]
such that for the quantum field $\langle\phi(x)\rangle = 0$, the energy density operator reads
${\hat\rho}(x)$ obtained from (\ref{eq:rho fluctuation}) is
\be {\hat\rho}(x) \equiv {\bar\rho}(t) + {\hat U}(x, t), \ee where ${\bar\rho}(t)= \half{\dot\psi}^2 +V(\psi)$ and ${\hat U}(x)$ involves 
the quantum field $\phi$. 

The formal structure of ${\hat U}(x,t)$, up to quadratic terms is written as
\be
{\hat U}(x) = r(t)\phi(x) + s(t) {\dot\phi}(x)+u(t){\phi^2}(x)+\half {\dot\phi^2}(x) +{1\over 2 a^2}(\nabla\phi)^2 \label{eq:quadratic}.
\ee

The coefficients $r(t),s(t)$ and $u(t)$ depend on the classical field and the parameters appearing in the potential $V(\psi)$.
In the potential driven inflation model. 

The inclusion of terms higher order in $\phi$ and its derivatives adds small nonlinearity and improves previous 
calculation to density fluctuation\cite{Kundu etal}.
The issue of renormalization of composite operators, determining short distance behaviour, as usual, not important in this context as 
measurements involve smearing of operator.

As the fluctuation size increases, a relativistic treatment is mandatory for superhorizon sized
perturbation. In contrast to density fluctuation for modes well within the horizon, contribution from the superhorizon modes 
are inflicted with the gauge ambiguity problem.
The evolution of perturbation in this regime is kinematic in nature involving the evolution of
curvature perturbation in space-time. A crucial result of such analysis is the constancy of a gauge
invariant quantity $\zeta$ for superhorizon sized modes \cite{Bardeen}. This is particularly useful in following the
density perturbation of a given mode from its exit from horizon during inflationary phase until its reentry in
radiation or matter dominated epoch at a later stage. In the uniform Hubble constant gauge, $\xi$
assumes a particularly simple form at the horizon crossings (${\it i.e.}~k/aH \sim 1$): \[ \xi = {\delta\rho\over \rho+p}~. \]
It may be noted that $\delta\rho$ is gauge non-invariant, thus \[ \zeta= (1+p/\rho)^{-1} \left(\delta\rho\over \rho\right)_H \]
is not manifestly gauge invariant.
	
\section{Density perturbation in power law inflation models}

 In this section we consider fluctuation in the energy density of scalar field in the background admitting power law inflation. Typically, we are interested in a scheme of calculating density perturbation spectrum that may be applicable to models such as extended inflation. Here we assume that, without much ado, the classical inflationary solution for the scalar field joins smoothly to the solutions before and after the inflation, which indeed is a subtle proposition\cite{Mallik-Leutwyler}. 
 
 On separating the field in classical and quantum parts, the energy density involving classical field (\ref{eq:rho fluctuation})	is
 \[ {\bar\rho}(t) =  \half{\dot\psi}^2 + V({\psi}),\] whereas the quantum part $U(x,t)$ is given by (\ref{eq:quadratic}).
 
Note that the correlation of a composite operator ${\cal{O}}(x)$ involving product of field and its derivatives is given by
 \[\langle {\cal{O}}(x){\cal{O}}({x^\prime})\rangle - \langle{\cal{O}}(x)\rangle.\langle{\cal{O}}({x^\prime})\rangle. \]
The correlation in energy density operator given in terms of the expectation values of field and its derivatives (\ref{eq:quadratic}) leading to
$$\langle U(\vec{x},t)U(\vec{x^\prime},{t^\prime})\rangle = \int {d^{3}k\over(2\pi)^3} e^{i\vec{k}.(\vec{x}-\vec{\acute{x}})}\tilde{P_U}(\vec{k}^{2};t,{t^\prime}), $$
where
\bqr
\tilde{P_U}(\vec{k}^{2};t,{t^\prime}) &=&[r(t)r({t^\prime})+s(t)s({t^\prime})\partial_{t}\partial_{{t^\prime}}
+\left\{r(t)s({t^\prime})\partial_{{t^\prime}} + (t\leftrightarrow{t^\prime})\right\}]\tilde{P}(\vec{k}^{2},t,{t^\prime})\nonumber\\
&+&{1\over2}\int {d^{3}k^\prime\over(2\pi)^3}\Big[4u(t)u({t^\prime})+{\partial_{t}}^{(1)}{\partial_{{t^\prime}}}^{(1)}{\partial_{t}}^{(2)}{\partial_{{t^\prime}}}^{(2)}
+\left\{{\vec{{k^\prime}}.(\vec{k}-\vec{{k^\prime}})}\over{a(t)a({t^\prime})}\right\}^{2}+\left\{u(t){\partial_{{t^\prime}}}^{(1)}{\partial_{{t^\prime}}}^{(2)}
+(t\leftrightarrow{t^\prime})\right\}
\nonumber\\&-&\vec{{k^\prime}}.(\vec{k}-\vec{{k^\prime}})\left\{{2u(t)\over{a({{t^\prime}})^{2}}}+{1\over{a^{2}(t)}}{{\partial_{{t^\prime}}}^{(1)}{{\partial_{{t^\prime}}}^{(2)}}}
+(t\leftrightarrow{t^\prime})\right\} \Big]\tilde{P}^{(1)}(\vec{{k^\prime}}^{2},t,{t^\prime})\tilde{P}^{(2)}((\vec{k}-\vec{{k^\prime}})^{2},t,{t^\prime}).
\label{eq:U correlation} 
\eqr
Here ${\tilde{P}}(k^{2};t,{t^\prime})$ is given by the expectation value of fields as
\bqr \langle\phi(x,t)\phi(x^\prime,t^\prime)\rangle = \int {d^{3}k\over(2\pi)^3} e^{i\vec{k}.(\vec{x}-\vec{\acute{x}})}\tilde{P}(\vec{k}^{2};t,{t^\prime}). \eqr

It may be mentioned that the highly divergent terms arising from composite operators are ignored here as measurement of a quantity at sharply defined
instant or at a precise point are hardly considered in quantum theory, rather it is perceived in a smeared sense\cite{Bartolo2,Banerjee-Mallik}. This is also sensible 
as the long wavelength limit is really of interest in the present context.

Here we parametrize the density matrix at the onset of inflation $t_0$ as $e^{-\beta_0 H(t_0)}$ leaving aside temporarily the question whether truly thermal equilibrium prevailed till before inflation and the reason for its departure thereof.\cite{Mallik-Leutwyler}

The two point function in real-time formulation \cite{Semenoff-Weiss2, Kundu etal} is given by

\begin{figure}
	\begin{center}
	\includegraphics[scale=0.25]{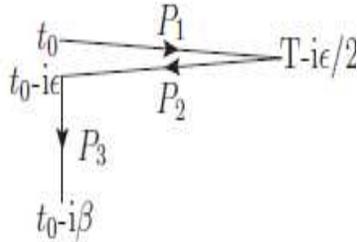}
		\caption{Closed time contour}
	\end{center}
\end{figure}

\be \tilde{P}(\vec{k}^{2};t,{t^\prime})={1\over{[a(t)a({t^\prime})]}^{3/2}}[g_k^{+}(t)g_k^{-}({t^\prime})+n(\omega_0)\left\{g_k^{+}(t)g_k^{-}({t^\prime})+g_k^{-}(t)g_k^{+}({t^\prime})\right\}],\label{eq:2 point function}\ee
where $n(\omega_0) =(e^{\beta_0\omega_0}-1)^{-1}$. 
where $\omega_0$ is related to the wavelength of the modes concerned, 
\bqr
 \omega_0^2(k)&=&{k^2\over a^2(t_0)}+H^2(t_0)\Big[12\xi \Big(1-\frac{2}{p}\Big) -\frac{9}{4}\Big(1-\frac{2}{3p}\Big)\Big]\nonumber\\
 &\equiv& 2\pi H(t_0)\Big[\Big({k\over 2\pi a(t_0)H(t_0)}\Big)^2+\gamma^2\Big]
 \eqr
and the mode functions $g^{\pm}$ are given in (\ref{eq:value of mode function}). 

Note that for the purpose of calculation of density perturbation the time at which mode function leaves the horizon (say, $t_H$) is important. 
So the concerned mode also expands by a factor $a(t_H)/a(t_0)$ during the time interval between onset of inflation $(t_0)$ and that of horizon exit $(t_H)$. 
Also the horizon size potentially sets the maximal limit on the scale of fluctuation to be considered \cite{Lyth}, and within it the oscillatory nature of mode functions
are clearly defined.

Using expression (\ref{eq:2 point function}), the time derivatives with respect to $t$ and $t^\prime$ can be calculated using power law evolution of scale factor 
and, in the equal time limit the first term in the expression (\ref{eq:U correlation}) is given by

\bqr
\tilde{P}_U(\vec{k}^2;t,t)&=& |r(t) + s(t) H(t)X(k,t)|^2 \tilde{P}(|\vec{k}|^2;t,t)\nonumber\\ 
&+& \half\int{d^3k^\prime\over (2\pi)^3} \Big|2u(t)+H^2(t) X(k^\prime,t)X(k-k^\prime,t) -{\big\{{\vec{k}^\prime}.(\vec{k}-\vec{{k}^\prime})\big\}\over a^2(t)}\Big|^2 \tilde{P}^{(1)}(|\vec{k}^\prime|^2;t,t)\tilde{P}^{(2)}(|\vec{k}-\vec{k}^\prime|^2;t,t),
\label{eq:2 point function 1}
\eqr
where, 
\[ X(k,t) = \left\{\half\left({1\over p}-3\right) - {k\over a(t)H(t)}{{H_\nu^{(1)}}^\prime (z)\over H_\nu^{(1)}(z)}\right\},\]
with
\be \nu=\sqrt{{(3p-1)^2\over{4(p-1)^2}}-6\xi{p(2p-1)\over{(p-1)^2}}}\quad
 {\rm{and}}\quad z= \{p/( p-1)\}.\{k/[a(t)H(t)]\}.\ee
 
Here terms that are suppressed by higher powers of $a(t)$, such as, 

 \bqr
 -2H(t)\left(u(t)-{\vec{{k^\prime}}.(\vec{k}-\vec{{k^\prime}})\over a^2(t)}\right)&.&
 \Big[{n(\omega_0(k-k^\prime))\over a^3(t)}.{k^\prime\over a(t)H(t)} {\rm Im}\left({{H_\nu^{(1)}}^\prime(z^\prime)\over H_\nu^{(1)}(z^\prime)}\right)\tilde{P}(\vec{k^\prime}^{2};t,{t^\prime})
\nonumber\\ &+&{n(\omega_0(k^\prime))\over a^3(t)}.{k-k^\prime\over a(t)H(t)}{\rm Im}
 \left({{H_\nu^{(1)}}^\prime(z-z^\prime)\over H_\nu^{(1)}(z-z^\prime)}\right)\tilde{P}^{(1)}(({\vec{k}}-\vec{k^\prime})^{2};t,t^\prime)\Big]\nonumber\\&-& {2\over a^6(t)}n(\omega_0(k^\prime))n(\omega_0(k-k^\prime)),~~\label{loop int}
 \eqr
are ignored in the term involving integration over $k^\prime$.

Note that in the equal time limit, 
\be
{k^3\over 2\pi^2}\tilde{P}(k^2;t,t) = \coth\Big({\beta_0\omega_0(k)\over 2}\Big).\pi^2H^2\left({p\over p-1}\right)\left({k\over 2\pi a(t)H(t)}\right)|H_\nu^{(1,2)}(z)|^2.
\ee

The first term in eqn.(\ref{eq:2 point function 1}) evaluated for the frozen modes at long wavelength scale using small argument approximation $k/aH << 1$ of Hankel function is

\be
{k^3\over 2\pi^2} \tilde{P}_1(k,t) = \coth\Big({\beta_0\omega_0(k)\over 2}\Big) \left|r(t)+s(t)H(t)\left\{(\nu-{3\over 2})-{1\over p}(\nu-{1\over 2})\right\}\right|^2{\Gamma^2(\nu)\over \pi^{2\nu}}\left({p\over {p-1}}\right)^{1-2\nu}\left({k\over 2\pi a(t)H(t)}\right)^{3-2\nu} H^2(t),
\ee

which agrees, in a limit, with conventional result using zero temperature field theory.

The second term in (\ref{eq:2 point function 1}) has the general structure

\be
	\half\int {d^3k^\prime\over (2\pi)^3} \left[A+B|{\vec k}^\prime|.|\vec{k}-{\vec k}^\prime|+C{\vec k}^\prime.(\vec{k}-{\vec k}^\prime)+ D\{{|{\vec k}|^\prime}^2, |\vec{k}-{\vec k}^\prime|^2,\vec{k}^\prime.(\vec{k}-{\vec k}^\prime)\}\right]P^{(1)}({\vec k}^\prime;t,t) P^{(2)}(\vec{k}-{\vec k}^\prime;t.t),  
	\label{eq. correlation structure}
\ee
where the coefficients are functions of parameters $u(t),H(t),a(t)$ and $p$. 

The integral over $k^\prime$ in eqn. \label{1 loop} is not constrained by the size of the horizon and it is difficult to evaluate exactly. However, one may estimate approximately the contributions from both the large sub-horizon and super-horizon size fluctuations separately.
In terms of scaled wavenumber $q = k/[2\pi a(t) H(t)]$, one has


\be
\tilde{P}^{(1)}(\vec{q^\prime})P^{(2)}(\vec{q}-\vec{q}^\prime)= \coth\left({\beta_0\omega_0(q^\prime)\over 2}\right)\coth\left({\beta_0
	\omega_0(q-q^\prime)\over 2}\right)2^{-4}\Big({p\over p-1}\Big)^{2}{\pi^2\over H^2a^6}|H_\nu^{(1)}(z_q)|^2|H_\nu^{(1)}(z_{q-q^\prime})|^2.
\ee

Using (\ref{eq. correlation structure}), the leading term can be estimated for small and large $k^\prime$, keeping in view that small $k$ is the region of interest, 
\bqr
{k^3\over 2\pi^2}\tilde{P}_U^{(int.)}(k,t) = A.{H^4\over 8}q^3\left({p\over p-1}\right)^{2-4\nu}\pi^{3-8\nu}
\Gamma^4(\nu)\int d^3q^\prime&&\Big[1+{2\over \beta_0}\Big({1\over \omega( q^\prime)}+{1\over\omega(|q-q^\prime|)}\Big)+\nonumber\\&&{4
	\over\beta_0^4}{1\over \omega(q^\prime)\omega(|q-q^\prime|)}\Big]|{\vec q}^\prime|^{-2\nu} |{\vec q}-{\vec q}^\prime|^{-2\nu}.
\label{small momentum}
\eqr
The integrals (\ref{small momentum}) can be evaluated using Feynman parametric integral representation, 
	\be
	{1\over A_1^{\nu_1} A_2^{\nu_2}\cdots A_n^{\nu_n}} = {\Gamma(\nu_1+\nu_2 +\cdots +\nu_n)\over \Gamma(\nu_1)\Gamma(\nu_2)\cdots \Gamma(\nu_n)} \int_0^1{d\alpha_1}\int_0^1{d\alpha_1}\int_0^1{d\alpha_2}\cdots\int_0^1{d\alpha_n}   {\alpha_1^{\nu_1 -1}\alpha_2^{\nu_2 -1}\cdots \alpha_n^{\nu_n -1}\delta(1-\Sigma \alpha_i) \over{[A_1\alpha_1+A_2\alpha_2+\cdots A_n\alpha_n]^{\nu_1+\nu_2+\cdots+\nu_n}}}
	\label{Feynman}
	\ee
	and dimensional regularization, with formally extending the range of integration fully, as an estimate, we have 
\be
A.H^4q^{6-4\nu}\left(p\over p-1\right)^{2-4\nu}\pi^{3-8\nu}\Gamma^2(\nu)
\Big[{\pi{\sqrt\pi}\over 8}
{\Gamma(2\nu -3/2)\Gamma^2({3\over 2}-\nu)\over \Gamma(3-2\nu)}+ {\pi\over 2}\left({a\over \beta_0 k}\right)\Gamma(2\nu -1) I_1 + {{\sqrt\pi}\over 2}\left({a\over \beta_0 k}\right)^2\Gamma(2\nu - 1/2)I_2\Big], 
\label{eq:density perturbation1}
\ee
where the integrals $I_1$ and $I_2$ are given by

\[
I_1 = \int_0^1 d\alpha_1\int_0
^{1-\alpha_1} d\alpha_2 {\alpha_1^{\nu -1}\alpha_2^{\nu -1}(1-\alpha_1-\alpha_2)^{1/2}\over\left[\alpha_1(1-\alpha_1)+(1-\alpha_1-\alpha_2)
	(\gamma^2/q^2)\right]^{2\nu+1/2}}
\]
and
\[
I_2 = \int_0^1 d\alpha_1\int_0
^{1-\alpha_1} d\alpha_2\int_0^{1-\alpha_1-\alpha_2} d\alpha_3\,{ \alpha_1^{-1/2}\alpha_2^{-1/2} \alpha_3^{\nu -1}(1-\alpha_1-\alpha_2-\alpha_3)^{\nu -1}\over\left[(\alpha_1+\alpha_3)(1-\alpha_1-\alpha_3)+(\alpha_1+\alpha_2)
	(\gamma^2/q^2)\right]^{2\nu+1}},
\] 
which are weakly dependent on $q$.
In particular the de Sitter limit $(\nu\rightarrow 3/2)$ can be obtained as
	
	\be
	A.{H^4\over 32\pi^5}\left[{\sqrt\pi}\left\{{1\over {3\over 2}-\nu}+\ln(\pi q^2)\right\}+{4\over{\sqrt \pi}}\left({a\over \beta_0k}\right) I_1+ 3\left({a\over \beta_0k}\right)^2I_2\right].
	\ee
Similarly for the large value of $k^\prime$ with small $k$, we have
\be
{k^3\over 2\pi^2}\tilde{P}_U^{(int.)}(k,t) = A.{H^4\over 8\pi}q^3\int d^3q^\prime \Big[1+{2\over \beta_0}\Big({1\over \omega(q)}+{1\over\omega(q-q^\prime)}\Big)+{4
	\over\beta_0^4}{1\over \omega(q)\omega(q-q^\prime)}\Big]{1\over|{\vec q}^\prime|}.{1\over|{\vec q}-{\vec q}^\prime|}
\ee

 The integral is independent of $\nu$ and it can be estimated                                                                                                                                                                                                                                                                             with range of $k^\prime$ extended fully and employing dimensional regularization 
 gives
 \be
 A. {\sqrt \pi}H^4q^4\left[-2+8\pi^2\Big\{-{1\over D-{3\over 2}}-2 \ln(4\pi q^2)-2\gamma +8\Big\}\Big({a\over\beta_0k}\Big)+ {\pi^3{\sqrt\pi}\over2}\Big({a\over\beta_0k}\Big)^2
 \right], 
 \ee 
 where $D\rightarrow 3/2$ and $\gamma$ is Euler-Mascheroni constant.

The fluctuation at the onset of inflation $t_0$ can be estimated using (\ref{eq:2 point function 1}) at time $t_0$ and can be extended until it exits the horizon at $t_H$. They are related as $q(t_0) =[a(t_H)/a(t_0)]\cdot [H(t_H)/ H(t_0)]q(t_H)$. After the exit from horizon,the perturbation length scale can be followed till horizon re-entry
following 

Further, from the beginning of inflation $t_0$ to the present epoch $t_p$, change in $q(t_0)$ is given by 
$$q(t_0) = {a(t_e)H(t_e)\over a(t_0)H(t_0)}\cdot {a(t_p)H(t_p)\over a(t_e)H(t_e)}\cdot q(t_p).$$ As the universe expanded adiabatically after inflation, the ratio given $a(t_p)/a(t_e)\sim T_e/T_p$, and the perturbation re-enters the horizon in present matter dominated stage is $H(t_p)/H(t_e) \sim t_e/t_p$.  Also the scale factor changes during inflation as $a(t_e)/a(t_0) =Z$, the amount of inflation undergone,
so 
\be
q(t_0)\sim Z\Big({t_0\over t_e}\Big)\Big({t_e\over t_p}\Big)\Big({T_e\over T_p}\Big) q(t_p)\sim Z\Big({t_0\over t_p}\Big)\Big({T_e\over T_p}\Big)q(t_p).
\ee
Considering present temperature $T_p=2.7K$ and typical perturbation size $k^{-1}$ between $(1-10^4)Mpc$
giving $${ k \over a(t_0)T(0)}\sim {Z\over 10^{25}}k_{Mpc}.$$
As most of the inflationary models require at least 70 e-folding of scale factor, making $k/[a(0)T(0)]$ large enough to consider $n[\omega_0(k)]=0$ for practical purposes\cite{}.

One may also note that the first term in (\ref{eq:2 point function 1}), in the limit
$p>>1$ i.e $\nu=3/2$, the de Sitter case, gives scale invariant spectrum with $|r(t_0)|$ related to derivative of potential (\ref{eq:2 point function 1}) 

\be 
{k^3\over 2\pi^2}P_U^{1}(k,t_0) \simeq
{H^2\over 8\pi}\Big({\partial V(\psi)\over\partial \psi}\Big)^2\Big|_{t_0}
\ee


From eqn (\ref{eq:density perturbation1}), the spectral index defined as $n_{s}-1 = d(ln \tilde{P}_U)/{d ln k}$ 
gives approximately by 
\be n_{s}-1\simeq (6-4\nu) \label{eq:spectral}\ee

One may note that the perturbation spectrum depends both on p and $\xi$ intricately. So one may constrain the parameter space in this quantities to fit with the observational results.

\begin{figure}[tbp]
	\includegraphics[scale=0.5]{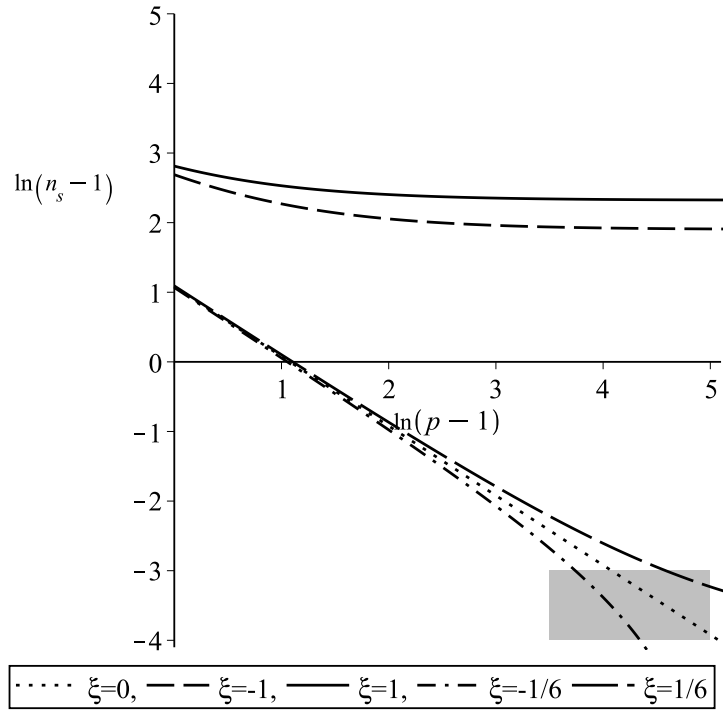}
	\caption{Distribution of spectral index in logarithmic scale for -3 $\leq \xi \leq$ 3}
	\label{fig:spectral}
\end{figure}

Equation~(\ref{eq:spectral}) shows the variation of $n_{s}$ with $p$ and coupling parameter $\xi$ in logarithmic plot figure~\ref{fig:spectral}. As evident from figure~\ref{fig:spectral}, the dependence of $n_{s}$ on $\xi$ is not negligible for high p values. The observational data \cite{WMAP,PLANCK}constrains the power law inflation within the range, 38 $\leq$ p $\leq$ 101 \cite{Sahni 3}. Given the range of spectral index $0.945 \leq n_s \leq 0.98 $, it is shown to be in agreement with the observational data \cite{WMAP,PLANCK} in the shaded region.

In particular, inclusion of quadratic terms in fluctuation gives rise to logarithmic dependence on wave number k in the de Sitter limit which is in resemblance with the usual results obtained such as in\cite{Bartolo2}.

\section{Conclusion}
 
Study of scalar field in time dependent cosmological background is important as it
has the potential to explain the observed density fluctuation in Universe from the quantum fluctuation originating in a small causal region at an early epoch.
We have addressed the issue of density perturbation within the scope of power law inflation in a simple model of scalar field coupled to scalar curvature in non-minimal fashion. Considering the infrared modes of the ultra-light scalar field, the average of energy-momentum tensor operator is fitted with perfect fluid model and a consistent, but infrared divergent, mode is shown to occur at $\xi<0$ region. As the density perturbation spectrum is related to the correlation of energy-momentum tensor, the leading part of the power spectra is determined in terms of two point function. The real time formalism of Semenoff-Weiss has been employed to obtain two point function. This can be followed reliably even to the time of leaving the horizon, depending upon modes. Although the formulation precisely depends on the time of onset of inflation but for practical purposes it may be insignificant. Inclusion of quadratic term in fields has been shown to lead to logarithmic correction in the de Sitter limit, usually found in the literature. The spectral index depends on $\xi$ and $p$, therefore relevant parameter space can be explored to check the power law inflation consistent with observational data.


\end{document}